\documentclass[12pt]{article}
\usepackage{epsfig}
\usepackage{ftnxtra}
\usepackage{hyperref}
\usepackage{graphicx}
\usepackage{natbib}

\def\beq{\begin{equation}}
\def\eeq#1{\label{#1}\end{equation}}
\def\eeqn{\end{equation}}

\def\beqa{\begin{eqnarray}}
\def\eeqa#1{\label{#1}\end{eqnarray}}
\def\eeqan{\end{eqnarray}}

\let\bar=\overbar

\def\Dslash{\not{\hbox{\kern-4pt $D$}}}
\def\dslash{\not{\hbox{\kern-2pt $\del$}}}

\def\msb{{\bar{\ssstyle M \kern -1pt S}}}

\def\Title#1{\begin{center} {\Large {\bf #1} } \end{center}}

\begin{document}

\Title{Optical interferometry and adaptive optics of bright transients}

\bigskip\bigskip

%+\addtocontents{toc}{{\it D. Reggiano}}
%+\label{ReggianoStart}

\begin{raggedright}  

{\it Florentin Millour, Olivier Chesneau, Anthony Meilland, Nicolas
  Nardetto\index{Millour, F.}\\
Lagrange Laboratory, 
UMR7293, Universit\'e de Nice Sophia-Antipolis, \\
CNRS, Observatoire de la C\^ote d’Azur, Bd. de l'Observatoire, 06304
Nice, FRANCE.\\
Send comments to \url{fmillour@oca.eu}}
\bigskip\bigskip
\end{raggedright}

\section{Introduction}

\begin{figure}[htbp]
  \begin{center}
    \includegraphics[width=0.7\textwidth]{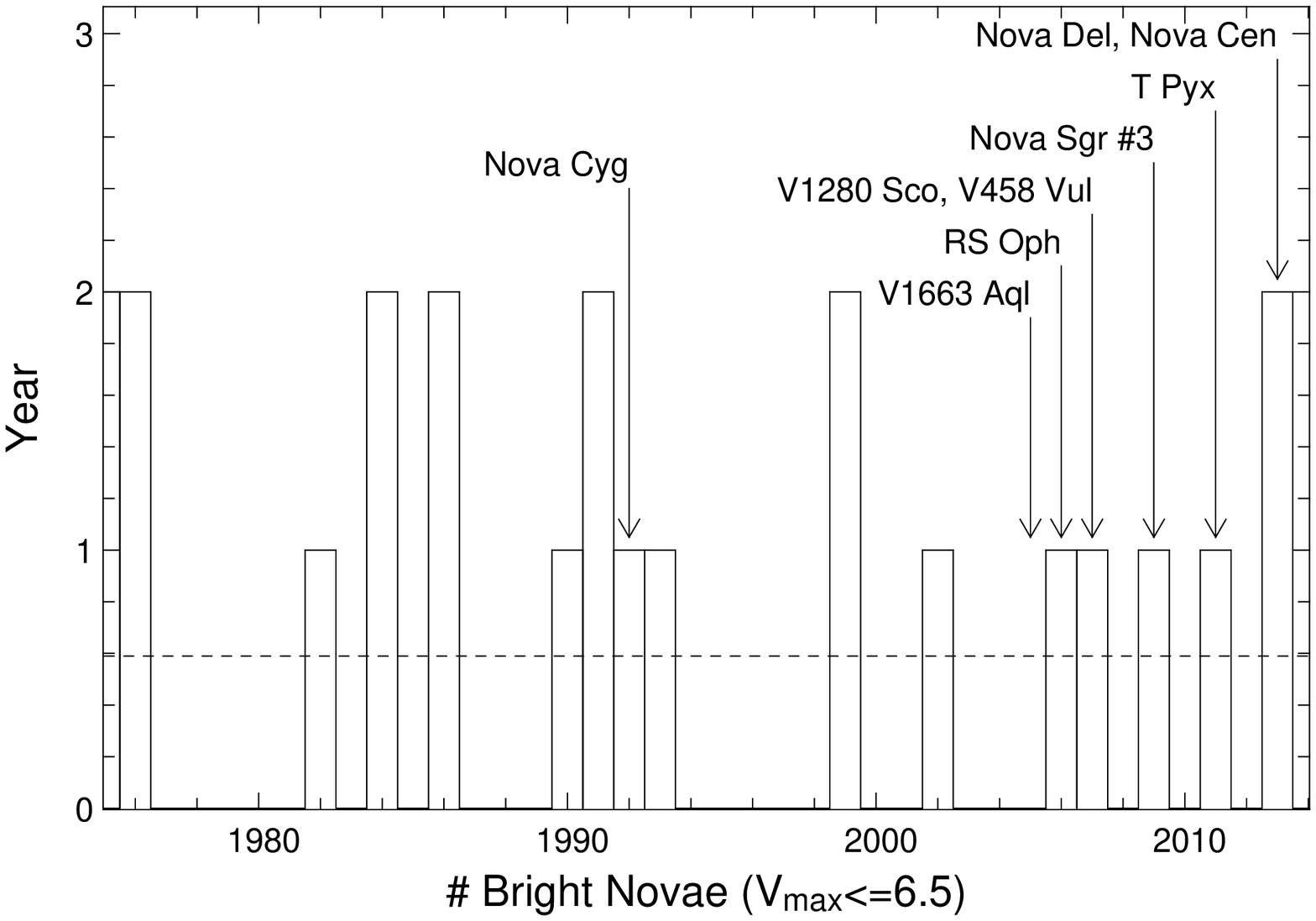}
    \caption{Typical frequency of bright nov\ae\  from 1975 to
      2013. About one nova every two years occurs, either in the North
      or South hemisphere\footnote{data from
        \url{http://www.cbat.eps.harvard.edu/nova_list.html},
        complemented by data from
        \url{http://asd.gsfc.nasa.gov/Koji.Mukai/novae/novae.html} .
        Some peak magnitudes were corrected using AAVSO data.}. V1663
      Aql \& V458 Vul do not appear in this plot as too faint in V
      (but bright in K). The average rate per year is plotted as a
      dashed line. The arrows show Nov\ae\  observed with optical
      interferometers.}
    \label{fig:novaeFreq}
  \end{center}
\end{figure}

Bright optical transients (i.e. transients typically visible with the
naked eye) are populated mainly by nov\ae\  eruptions plus a few
supernov\ae\  (among which the SN1987a event). Indeed, usually one bright
nova happen every two year, either in the North or the South
hemisphere (see Fig.~\ref{fig:novaeFreq}). It occurs that current
interferometers have matching sensitivities, with typically visible or
infrared limiting magnitudes in the range 5--7. The temporal
development of the fireball, followed by a dust formation phase or the
appearance of many coronal lines can be studied with the VLTI. The
detailed geometry of the first phases of nov\ae\  in outburst remains
virtually unexplored. This paper summarizes the work which has been
done to date using mainly the Very Large Telescope Interferometer.

We invite the reader to have a look to the extensive review on the
topic by \citet{Chesneau2012a} for a complete description of the
science on transients that can be achieved with optical/infrared
long-baseline interferometers. We give a short summary of the content
of this paper in the next section.

\section{Why observing nov\ae\  with optical interferometers?}

Optical interferometers represent a breakthrough in terms of spatial
resolution, that can provide crucial information related to the nova
phenomenon. All targets in the 3-5~kpc range can be potentially
resolved by current interferometers (CHARA, VLTI, NPOI).

The VLT Interferometer can provide measurements of the angular
diameters of the nova ejecta in continuum and lines, from the near-IR
to the mid-IR in the very first moments of the outburst. The primary
outcome of these observations is a direct estimate of the expansion
parallax, thus the distance to nov\ae. Of importance is the possibility
to spatially and spectrally resolve different near-IR emission lines
to estimate the physical conditions throughout the wind and the
ejecta. For those objectives, the medium spectral resolution of the
near-IR instrument AMBER is an asset. If the nova appears to form dust
(CO nov\ae\ ), an in-depth study of the dust forming regions can be
carried out with the MIDI instrument. Using a set of flexible
observing runs, we shall follow the outburst from the first days up to
several months.

%%%%%%%%%%%%%%%%%%%%%%%%%%%%%%%%%%%%%%%%%%%%%%%%%%%%%%%%%%%%%%%%%%%%%%%%% 
%% 
%% use this format to include an .eps figure into your paper
%% 

%%%%%%%%%%%%%%%%%%%%%%%%%%%%%%%%%%%%%%%%%%%%%%%%%%%%%%%%%%%%%%%%%%%%%%%%%%% 

\section{Nova as a spherical fireball}

\begin{figure}[htbp]
  \begin{center}
    \includegraphics[width=0.7\textwidth]{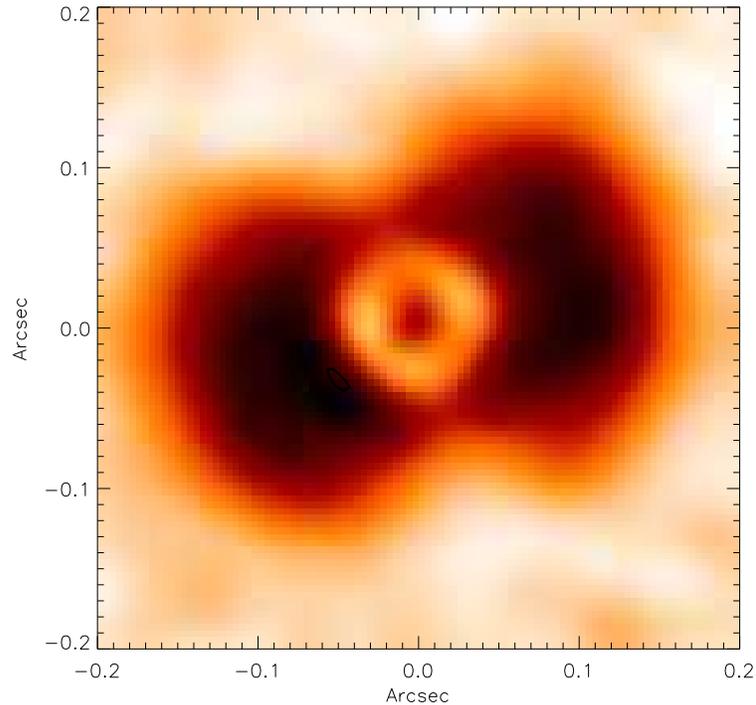}
    \caption{2010 NACO K band image after a PSF subtraction revealing
      the impressive bipolar nebula. Mid-IR images also show that
      there is no dust emission in the equatorial plane.}
    \label{fig:NACO}
  \end{center}
\end{figure}

Up to now, the program of observations has focused on nov\ae\  with
magnitudes reachable by the VLTI (South declination \& K magnitude
$\geq 7$). Past observations and theoretical work on the nova
phenomenon have provided a substantial knowledge about the physical
nature of these binary systems and the outburst. However, these
investigations are naturally limited by the difficulty of estimating
the distance, which is usually inferred indirectly and with large
errors. Spherical symmetry is a basic tenant adopted in the
derivation of relationships that link the non-spatially resolved
photometric and spectroscopic observations to the physical parameters
of the system \citep[][Table 3]{Gehrz1998a}. Spherical symmetry is
implicitly assumed when the $uv$ coverage is not sufficient to perform
a better analysis. This was the case for the Nova V1280\,Sco
\citep{2008A&A...487..223C} which was also observed exclusively using
2 telescope recombination. Two years after, observations with the AO
system NACO mounted at the UT4 telescope revealed an impressive dusty
bipolar nebula \citep[see Fig.~\ref{fig:NACO}]{Chesneau2012}.

%%%%%%%%%%%%%%%%%%%%%%%%%%%%%%%%%%%%%%%%%%%%%%%%%%%%%%%%%%%%%%%%%%%%%%%%% 
%% 
%% use this format to include an .eps figure into your paper
%% 

%%%%%%%%%%%%%%%%%%%%%%%%%%%%%%%%%%%%%%%%%%%%%%%%%%%%%%%%%%%%%%%%%%%%%%%%%%% 

\section{A Bipolar fireball from the first blink}

An interferometer is mainly sensitive to the angular size of a nova in
its early stages. Measuring the size of the fireball in different
orientations on-sky allows us to infer the axis ratio and orientation
of an individual nova shell. This is relatively easy to obtain for a
3--6-telescopes interferometer. This was the case for the outburst of
the recurrent nova RS\,Oph \citep{2007A&A...464..119C}.

The highly collimated outflow from the RS Ophiuchi has been imaged by
the HST \citep{Bode2007e} and in the radio
\citep{2006Natur.442..279O}. The AMBER observations showed that the jet
was already existing 5.5 days after the discovery, and provided a
unique view of radial-velocities which could afterwards complement the
expansion rates derived by the HST and radio images
\citep{2009ApJ...703.1955R}.

The signature of a bipolar jet in interferometric data is now well
identified provided that the emission lines are spectrally resolved
(R$\sim$1500). The famous nova T\,Pyx exhibited a spherical appearance
in broadband PIONIER data, but the signature of a bipolar kinematics
was clearly detected in our spectrally resolved AMBER data
\citep{2011A&A...534L..11C}. The numerous peculiarities of the T\,Pyx
eruptions can be explained in the frame of recurrent nearly face-on
eruptions that launch fast material in the line-of-sight and slow
material perpendicular to it, building up the slow expansion shell
imaged by the HST.

\section{Intermediate Luminosity Optical Transients}

Intermediate-Luminosity Optical Transients (ILOTs), are eruptive stars
with peak luminosity between those of nov\ae\  and supernov\ae\  that have
been also called Red Nov\ae\  or red Transient. The powering processes
and whether they are due to binary interaction or are formed through
single star evolution are debated. High angular resolution techniques
can play a role by tracking bipolarity and the formation of
disks. Furthermore, one can also study the remaining central star when
a merger is highly suspected, for instance by detecting the
deformation due to very high rotational rate.

\subsection{Sakurai's object}

In 1996, Sakurai's object (V4334 Sgr) suddenly brightened in the
center of a faint Planetary Nebula (PN). This very rare event was
interpreted as being the reignition of a hot white dwarf that caused a
rapid evolution back to the cool giant phase. From 1998 on, a copious
amount of dust has formed continuously, screening out the star that
remained embedded in this expanding high optical-depth
envelope. Mid-IR interferometry performed in 2008 with the MIDI/VLTI
instrument discovered a unexpectedly compact (30 × 40 milli-arc-second, 105
× 140 AU assuming a distance of 3.5 kpc), highly inclined, dust disk
\citep{2009A&A...493L..17C}. The major axis of the disk is aligned with
an asymmetry seen in the old PN. This implies that the mechanism
responsible for shaping the dust envelope surrounding Sakurai's object
was already at work when the old PN formed, a strong argument for
binary interaction.

%%%%%%%%%%%%%%%%%%%%%%%%%%%%%%%%%%%%%%%%%%%%%%%%%%%%%%%%%%%%%%%%%%%%%%%%% 
%% 
%% use this format to include an .eps figure into your paper
%% 
% \begin{figure}[htbp]
%   \begin{center}
%     \includegraphics[width=0.3\textwidth,angle=-0,origin=b]{PRODUCT_V838_Mon_1.46-1.86micron_2013-04-15T00_20_32.7510_UVplane.ps}\hspace{0.7cm}
%     \includegraphics[width=0.5\textwidth,angle=-90,origin=b]{V838_Mon_Fit.ps}
%     \caption{Left: UV plane obtained on V838 Mon in 2013 with
%       AMBER. Right: Data plotted as a function of spatial frequencies
%       and best-fit uniform disk model (black line, diameter 3\,mas). The lower dashed
%       line show the model for the size upper limit 4.7\,mas derived
%       from the data.}
%     \label{fig:V838Mon}
%   \end{center}
% \end{figure}
%%%%%%%%%%%%%%%%%%%%%%%%%%%%%%%%%%%%%%%%%%%%%%%%%%%%%%%%%%%%%%%%%%%%%%%%%%% 

\subsection{V 838 Mon}

V838 Monocerotis erupted in 2002, brightening by 9 magnitudes in a
series of outbursts, and eventually developing a spectacular light
echo.  A very red star emerged surrounded by copious amount of new
dust that condensed from the expanding ejecta of the outbursts. V838
Mon is the close-by archetype of the ILOT sources which are triggering
very active research currently.  MIDI/VLTI observations obtained over
the last few months showed that the dust resides in the form of a
flattened structure ($\sim$15-50 mas from 8 to 13 $\mu$m), i.e. a
90x300\,AU flattened structure for a distance of 6.2\,kpc (Sparks et
al. 2008). The modelling of this extended structure is in progress but
it is incomplete without a much better knowledge of the central source
which is seen as a very cool M-L type super-giant. AMBER observations
were also obtained in 2013 to measure the size of the central source,
its shape (since it has potential to be a fast rotator) and study the
cool photosphere / dusty disk transition. The AMBER data
%(Fig.~\ref{fig:V838Mon})
were essentially acquired with small baselines ($\leq30$\,m) and are
quite noisy ($\sigma V^2\approx0.05$), leading to an angular size of
3\,mas, but with a large uncertainty. We can basically only set an
upper limit to the diameter of the HK-bands object, of 4.7\,mas. This
would make the HK-bands object smaller than 30\,AU. This size (and
shape) difference between the HK-bands and N band is striking and
reminiscent of super-giant stars dusty disks \citep{Millour2011,
  Wheelwright2012}. Further AMBER observations would enable us to
pinpoint more precise properties of this intriguing object.

\section{Conclusion and prospectives}

We presented here a few results from the VLTI campaigns on Nov\ae\  and
ILOTs. These campaigns present a challenge in terms of scheduling and
observatory response but provide unique insights on the early
processes at stake when a nova explode. With the upcoming infrared
instruments like MATISSE \citep{2006SPIE.6268E..31L} or GRAVITY
\citep{2005AN....326..561E}, getting a finer idea of the geometry will
be much faster than with current instruments. The development of
visible interferometric instruments at the CHARA array or at the VLTI
would also be an asset to get the sharpest multi-wavelength picture of
these objects a few days after outburst before the advent of the ELT
in the mid-2030s which will enable direct snapshot pictures of the
fireball at 10's-milli-arc-seconds resolution for a much larger number
of nov\ae.

%______________________________________________________________
\bibliographystyle{aa} % style aa.bst 
\bibliography{biblioHWTU3} % your references Yourfile.bib

\end{document}